\documentclass[aps,prb,twocolumn,showpacs,preprintnumbers,superscriptaddress,amsmath,amssymb,longbibliography]{revtex4-2}
\usepackage{graphicx}% Include figure files
\usepackage[normalem]{ulem}
\usepackage[svgnames]{xcolor}
\usepackage{bm}% bold math
\usepackage{multirow}\usepackage{float}
\usepackage{titlesec}
\usepackage{physics}
\usepackage{xcolor}
\usepackage{subfigure}
\usepackage{subfig}
\usepackage{lipsum, babel}
\usepackage[font=small,labelfont=bf,
   justification=justified,
   format=plain]{caption}  
\usepackage{multirow}
\usepackage{soul}

\begin{document}
\title{Role of native point defects and Hg impurities in the electronic properties of Bi$_4$I$_4$}

\author{Gustavo H. Cassemiro}
\affiliation{Departamento de F\'isica, Universidade Federal de Minas Gerais,  C. P. 702, 30123-970, Belo Horizonte, MG, Brazil}
\author{C. David Hinostroza}
\affiliation{Laboratory for Quantum Matter under Extreme Conditions,  Instituto de Física, Universidade de São Paulo, 05508-090, São Paulo, Brazil}
\author{Leandro Rodrigues de Faria}
\affiliation{Escola de Engenharia de Lorena - DEMAR, Universidade de Sao Paulo, 12612-550, Lorena, Brazil}
\author{Daniel A. Mayoh}
\affiliation{Department of Physics, University of Warwick, Coventry CV4 7AL, United Kingdom}
\author{Maria C. O. Aguiar}
\affiliation{Departamento de F\'isica, Universidade Federal de Minas Gerais,  C. P. 702, 30123-970, Belo Horizonte, MG, Brazil}
\author{Martin R. Lees}
\affiliation{Department of Physics, University of Warwick, Coventry CV4 7AL, United Kingdom}
\author{Geetha Balakrishnan}
\affiliation{Department of Physics, University of Warwick, Coventry CV4 7AL, United Kingdom}
\author{J. Larrea Jim\'enez}
\affiliation{Laboratory for Quantum Matter under Extreme Conditions,  Instituto de Física, Universidade de São Paulo, 05508-090, São Paulo, Brazil}
\author{Antonio Jefferson da Silva Machado}
\affiliation{Escola de Engenharia de Lorena - DEMAR, Universidade de Sao Paulo, 12612-550, Lorena, Brazil}
\author{Valentina Martelli}
\affiliation{Laboratory for Quantum Matter under Extreme Conditions,  Instituto de Física, Universidade de São Paulo, 05508-090, São Paulo, Brazil}
\author{Walber H. Brito}
\affiliation{Departamento de F\'isica, Universidade Federal de Minas Gerais,  C. P. 702, 30123-970, Belo Horizonte, MG, Brazil}

\thanks{}

\date{\today} 
\begin{abstract}

We studied the effects of point defects and Hg impurities in the electronic properties of bismuth iodide (Bi$_4$I$_4$). Our transport measurements after annealing at different temperatures show that the resistivity of Bi$_4$I$_4$ depends on its thermal history, suggesting that the formation of native defects and impurities can shape the temperature dependence of electrical resistivity. Our density functional theory calculations indicate that the bismuth and iodine antisites, and bismuth vacancies are the dominant native point defects. We find that bismuth antisites introduce resonant states in the band-edges, while iodine antisites and bismuth vacancies lead to a $n$-type and $p$-type doping of Bi$_4$I$_4$, respectively. The Hg impurities are likely to be found at Bi substitutional sites, giving rise to the $p$-type doping of Bi$_4$I$_4$. Overall, our findings indicate that the presence of native point defects and impurities can significantly modify the electronic properties, and, thus, impact the resistivity profile of Bi$_4$I$_4$ due to modifications in the amount and type of carriers, and the associated defect(impurity) scattering. Our results suggest possible routes for pursuing fine-tuning of the electronic properties of quasi-one-dimensional quantum materials.
\end{abstract}
\maketitle

\section{Introduction}

The quasi-one dimensional bismuth halogenides have attracted the attention of the scientific community due to their emerging quantum properties, such as superconductivity and non-trivial topology~\cite{han_review_bi4x4}. Among this family of compounds,  bismuth iodide (Bi$_4$I$_4$) and bismuth bromide (Bi$_4$Br$_4$) were proposed to host weak, strong, or high-order topological insulating phases as well as pressure-induced superconductivity, depending on the crystal structure \cite{autes_novel_2016, Noguchi2021, huang2021room}.

In both Bi$_4$I$_4$ and Bi$_4$Br$_4$, the molecular chains can stack in two distinct forms, giving rise to the so-called $\alpha$ and $\beta$ phases. More interestingly, these compounds can be viewed as controlled platforms where one can go from one phase to another by tuning the temperature. In the case of Bi$_4$I$_4$, this temperature-driven first-order phase transition takes place around $T_C \approx 300$ K~\cite{huang2021room}. 
For $T > 300$ K, Bi$_4$I$_4$ crystallizes in the $\beta$ phase, which has been reported to host a non-trivial topological phase.
Density functional theory (DFT) and GW calculations have indicated that the $\beta$ phase is a weak topological insulator~\cite{liu2016weak,autes_novel_2016}, which was validated by angle-resolved photoemission spectroscopy~\cite{autes_novel_2016}. Regarding the $\alpha$ phase, which occurs for $T < 300$ K, it was earlier reported as a trivial insulator according to the $Z_2$ criterion. However, more recent studies found evidence of its high-order topological character~\cite{Noguchi2021}, dictated by low-energy electronic states originating mostly from the bismuth $p$-orbitals~\cite{liu2016weak}.

The complex interplay between the topological phases and superconductivity was explored in bismuth halogenides using external pressure as a tuning parameter. For Bi$_4$Br$_4$, it was found that the non-trivial topological phase is robust at higher pressures, indicating a possible coexistence between superconductivity and topology~\cite{Li_PNAS2019}.
Moreover, in the work of Qi and coworkers~\cite{Qi2018}, it was found that Bi$_4$I$_4$ undergoes multiple topological phase transitions before entering into the superconducting phase. In addition, it was found that Bi$_4$I$_4$ undergoes several structural transitions with the increasing of pressure, from monoclinic ($C2/m$) to trigonal ($P31c$), and then from the trigonal to a tetragonal ($P4/mmm$) structure~\cite{Deng_PRB2019}. 

The investigation of quantum oscillations in electrical transport pointed out the existence of 2D or 3D Fermi surfaces in $\alpha$-Bi$_4$I$_4$ for $n$ and $p$-doped samples, respectively~\cite{PRM_Chen2018}. A 3D Fermi surface was also revealed for the $\beta$-Bi$_4$I$_4$, for magnetic fields below 15.9 T, where there is a near metal to insulator transition~\cite{PRB_Wang2021}. Although those works indicated that defects and impurities might have an important role, their impact on the electronic structure has not been reported yet. Mu \textit{et al.}~\cite{Mu2023} reported that complex hollow-type defects, which appear in the regions of $\alpha$ and $\beta$ phase separation, modulate the density of states of Bi$_4$I$_4$, which might explain the sample-dependence observed in the resistivity measurements. In our recent work, we pointed out that simple native defects can be the source of the instability of the $\beta$ phase at low temperatures after quenching~\cite{David_Bi4I4}, suggesting that it is not the different topology of the phase that rules the observed temperature evolution of the resistivity, shedding some light on the source of the rather confusing resistivity measurements reported so far for the two phases when it come to the comparison of the temperature-dependence that does not seem to be ruled by the different topological phase.

Although Bi$_4$I$_4$ has been intensively studied in the last few years, there is a lack of a systematic study about the effects of native point defects and impurities on the electronic properties of both $\alpha$ and $\beta$ phases. We emphasize that scrutinizing the presence of different defects, which are intrinsic to all samples, may help to understand the wide variety of resistivity profiles which are impacted by intrinsic and extrinsic doping and the appearance of defect/impurity electronic states, and to avoid misleading interpretation related to the expected electronic properties stemming from a specific topological phase. In this work, we focus on the effects of native defects and Hg impurities on the electronic properties of both $\alpha$ and $\beta$ phases of Bi$_4$I$_4$. In particular, we have addressed the formation energies and electronic structure of native point defects and mercury (Hg) impurities in Bi$_4$I$_4$. Hg impurities were considered due to possible contamination of samples during the Bi$_4$I$_4$ synthesis. The presence of Hg is experimentally observed on the crystal surface of our samples by energy-dispersive x-ray spectroscopy (EDX) investigation.

We find that bismuth antisites are the dominant native defects, followed by iodine antisites and bismuth vacancies. More importantly, we observe that bismuth antisites do not dope the system, while iodine antisites and vacancies lead to $n$ and $p$-type doping of Bi$_4$I$_4$, respectively. The Hg antisites act like $p$ dopants. Overall, our calculations show that energetically stable point defects and Hg impurities modify the amount of carriers in Bi$_4$I$_4$, introduce additional scattering states around the Fermi energy, and may have significant effects on the resistivity of Bi$_4$I$_4$ at low-temperatures. Thus, the analysis of the features observed in electrical transport experiments of Bi$_4$I$_4$ as a consequence of a specific topological phase should consider these findings to avoid misleading interpretations.

Our paper is organized in three sections.  In Sec.~\ref{sec:method}, we present the employed experimental and theoretical methods. The EDX and resistivity measurements are discussed in Sec.~\ref{subsec:edx_resistivity}. The formation energies of the investigated defects and impurities are discussed in Sec.~\ref{subsec:form_defectsq0} and ~\ref{subsec:form_impuritiesq0}, and their effects on the electronic structure are discussed in Sec.~\ref{subsec:elec_defectsq0}. Our conclusions are summarized in Sec.~\ref{sec:conc}.

\section{Experimental and Computational Methods} 
\label{sec:method}

$\alpha$-Bi$_4$I$_4$ single crystals were grown by chemical vapor transport as explained in Ref.~\cite{David_Bi4I4}. For the thermal treatment samples were annealed for 24 hrs at 50~$^\circ$C and 100~$^\circ$C, and cooled slowly to smoothly recover the $\alpha$ phase, in contrast to what we reported in \cite{David_Bi4I4} where a rapid quenching aimed at stabilizing a metastable $\beta$-phase at low temperature. 
EDX was used on a scanning electron microscope (SEM) for the compositional analysis of the synthesized crystals to check the stoichiometry and the presence of other atomic elements. Electrical resistivity measurements were performed using an Advanced Research System DE-202N
cryocooler and a Quantum Design - Physical Property Measurement System in a 4-point configuration.  Electrical contacts on the samples were made using gold wires fixed with silver paint on Pt electrodes previously deposited on the surface to minimize the contact resistance.

Our first-principles calculations were performed within the density functional theory as implemented in the Vienna \textit{ab initio} simulation package (VASP)~\cite{vasp1,vasp2}. The Perdew-Burke-Ernzehof generalized gradient approximation (PBE-GGA) was used to perform the calculations. However, this functional cannot describe  the van der Waals interactions presented in Bi$_4$I$_4$, so the dispersion correction (vdW-DF)~\cite{vdw_DF} was employed. The spin-polarized and spin-orbit coupling (SOC) were also considered in our calculations. In addition, we employed projector augmented wave (PAW)~\cite{blochl1994projector} potentials, and we relaxed the structures until the forces on each atom were $<0.01$ eV/\AA\  and total energies converged within a $1\times 10^{-6}$ eV criterion. We used an energy cutoff of 400 eV for the plane-wave basis. The total energies were calculated using a \textit{k}-point mesh of $4 \times 3 \times 4$ for the $\alpha$ phase and $4 \times 3 \times 2$ for the $\beta$ phase. The supercell approximation was employed to investigate the energetics and electronic properties of Bi$_4$I$_4$ in the presence of native defects and Hg impurities. 

The relative equilibrium concentration of defects in the system depends on their formation energies, which can be obtained from the DFT total energies as follows~\cite{freysoldt2014first,komsa2012finite},
\begin{equation}
    \label{Eq:defect_form_energy}
    \begin{split}
    &E^{f}(S_{L}) = [E_{\text{tot}}(S_{L}) - E_{\text{tot}}(\text{pristine})] - \sum_i n_i\mu_i,    
    \end{split}
\end{equation}
where $E_{\text{tot}}(S_{L})$ and $E_{\text{tot}}(\text{pristine})$ are the total energies of supercells with the presence of defect $S_L$ (Kröger-Vink notation) and without the presence of defects, respectively. $n_i$ is the number of species $i$ that is added to ($n_i>0$) or removed from ($n_i<0$) the supercell to form the defect $S_L$. $\mu_i$ are the chemical potentials of different atomic species exchanged with their respective atomic reservoirs, obtained with reference to the calculated reference energies $\mu_i^0$, so  $\mu_i=\mu_i^0+\Delta \mu_i$. The values of $\Delta \mu_i$ depend on the environmental conditions in thermodynamic equilibrium. Due to the small band gaps of $\alpha$ and $\beta$ phases we considered only neutral charge defects and impurities~\cite{autes_novel_2016, liu2016weak}.

\section{Results and Discussions}

\subsection{EDX and resistivity measurements}
\label{subsec:edx_resistivity}

\begin{figure}
    \centering
    \includegraphics[width=7.5cm]{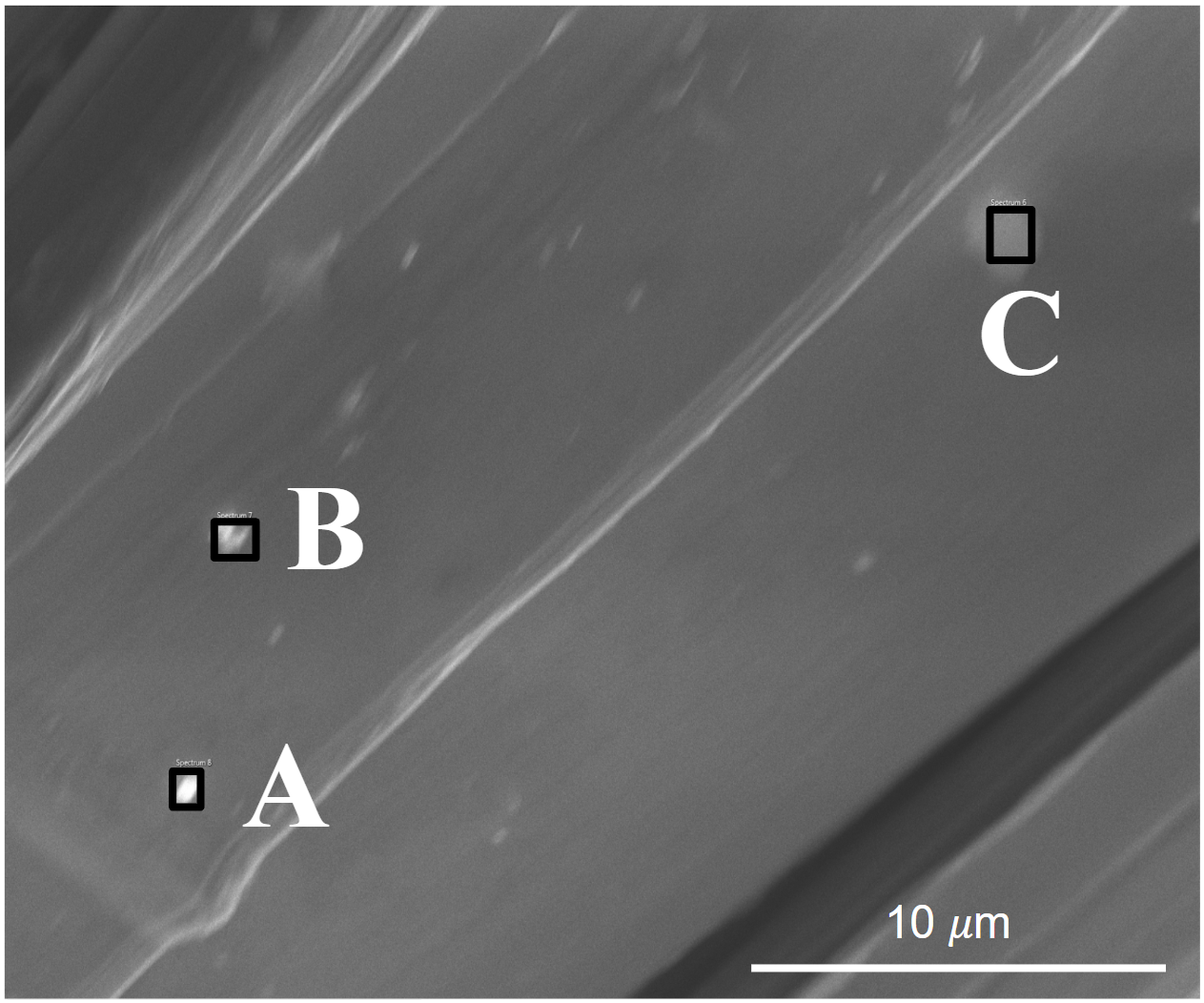}
    \caption{SEM image of one representative Bi$_4$I$_4$ sample. Chemical composition was measured in the areas enclosed by numbered squares by EDX. The elemental composition is given in Table~\ref{tab:EDX_com}.}
    \label{fig:SEM_HgBatch}
\end{figure}

In Fig.~\ref{fig:SEM_HgBatch} we show the SEM image of one representative Bi$_4$I$_4$ sample.
Compositional analysis has been done by EDX analysis across selected areas of the single crystal shown in Fig.~\ref{fig:SEM_HgBatch} and are presented in Table \ref{tab:EDX_com}. The observed presence of Hg impurities on the surface and the similar ionic radius of Bi (117 pm) and Hg (114 pm) \cite{shannon1976revised} suggest the possibility of Hg substitution in the bismuth site, although the evidence is not conclusive.

\begin{figure}[!ht]
    \centering
    \includegraphics[width=0.9\linewidth]{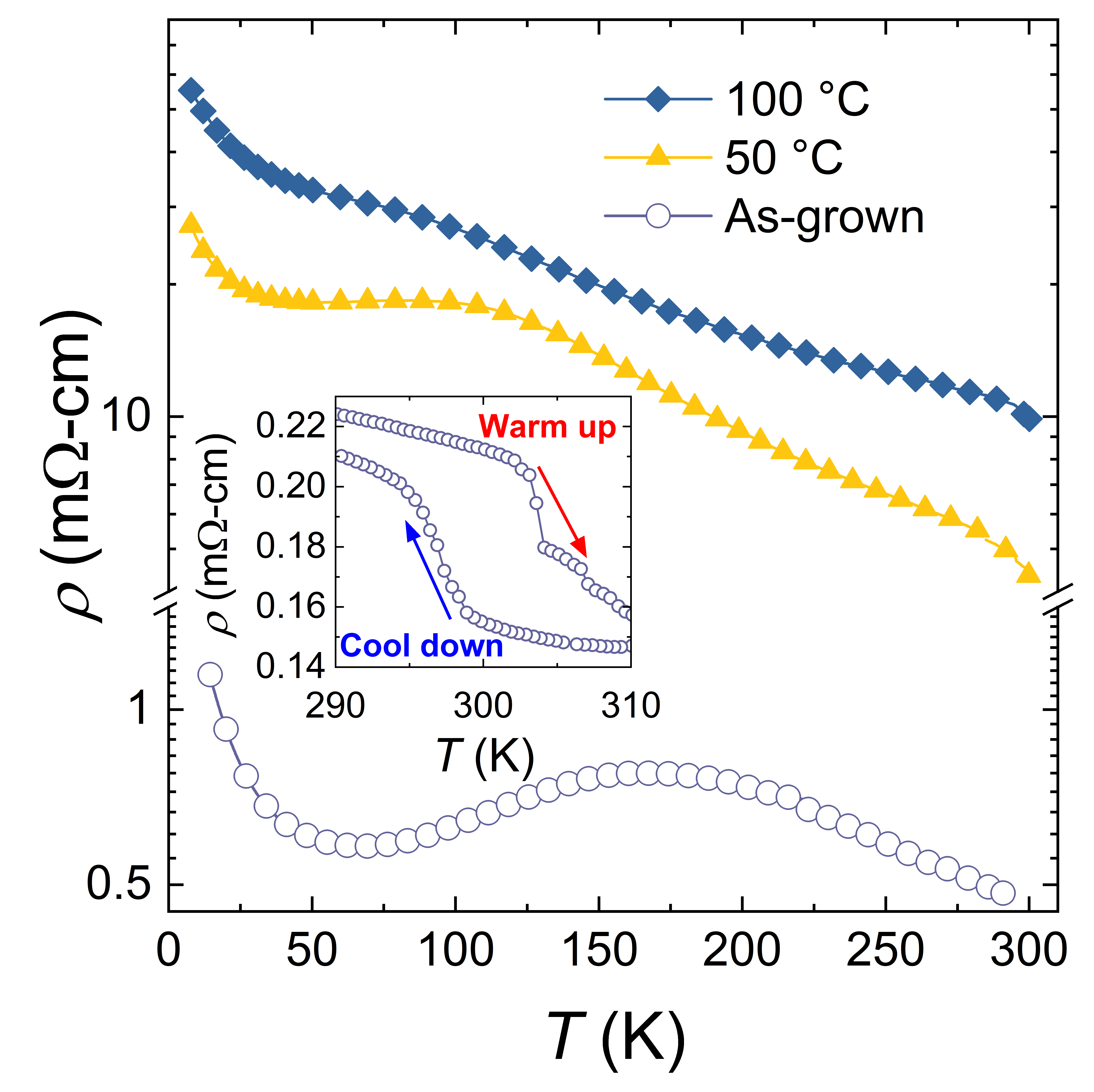}
    \caption{Temperature dependence of the electrical resistivity of $\alpha$-Bi$_4$I$_4$ samples. Filled symbols correspond to samples annealed for one day, while open circles to the as-grown sample. The inset shows the hysteresis curve, indicating the phase transition.}
    \label{fig:RhovsT}
\end{figure}

The temperature dependence of electrical resistivity on selected samples of Bi$_4$I$_4$ after annealing at different temperatures is depicted in Fig.~\ref{fig:RhovsT}. In our previous work~\cite{David_Bi4I4}, we scrutinized the large variety of reported electrical resistivity of this compound. We observed that the temperature dependence does not seem to correlate with the topological (structural) phase. However, a direct comparison among the available data was not possible, as the growth processes, annealing, and quenching temperatures vary across the reported works. 
Here, we systematically investigated the temperature dependence of the electrical resistivity of Bi$_4$I$_4$ samples from the same batch, just annealing them at different temperatures. We confirm (see inset of Fig.~\ref{fig:RhovsT}) that there is the expected hysteresis across the structural transition temperature. 
A smooth crossing through $T_C$ avoids phase coexistence for higher ($T>310$~K) and lower temperatures ($T<290$~K). Additionally, Fig.~\ref{fig:RhovsT} shows an clear change of both the absolute value and of the temperature dependence of electrical resistivity in the $\alpha$-phase after the annealing process; electrical resistivity increases by over one order of magnitude at room temperature after annealing. This feature suggests that the change in charge carrier density and additional scattering associated with the formation of native defects and impurities due to the samples' thermal history can likely be the source of the reduced conductivity. This evidence suggests that native defects and impurities have an important role: their presence can shape electronic properties within the same topological phase and will be analyzed and discussed in the following sections.

\begin{table}[!tb]
	\centering
	\caption{Composition measured by EDX across areas indicated in Fig.~\ref{fig:SEM_HgBatch}. The expected nominal composition of the single crystals is 50:50 for Bi:I.}
	\label{tab:EDX_com}
   \begin{ruledtabular}
	\begin{tabular}{cccc}
		\textbf{spectrum label} &  \textbf{I (\%)} & \textbf{Bi (\%)} & \textbf{Hg (\%)}\\ \hline 
    A & 66.49(3) & 33.51(2) & - \\
    B & 66.61(3) & 34.39(2) & - \\
    C & 49.14(2) & 46.61(2) & 4.25(2) \\
       
     %%\bottomrule
	\end{tabular}
    \end{ruledtabular}
\end{table}

\subsection{Defect-formation energies under distinct growth conditions}
\label{subsec:form_defectsq0}

We first investigate the energetics of different types of native defects which may appear in Bi$_4$I$_4$ under Bi-rich and poor conditions. In particular, we have considered bismuth (Bi$_\text{I}$) and iodine (I$_{Bi}$) antisites, bismuth (Bi$_i$) and iodine (I$_i$) interstitials, and finally bismuth (V$_{\text{Bi}}$) and iodine (V$_\text{I}$) vacancies. We have considered all the possible interstitial and substitutional sites, labeled as A, B, and C in the $\alpha$ phase;  A', B', C', D' in the $\beta$ phase. The defects were introduced in ideal $3 \times 1 \times 2$ (96 atoms) and $3 \times 2 \times 3$ (144 atoms) supercells for $\alpha$ and $\beta$ phases, respectively. In our previous work, we only obtained the formation energies of these defects under Bi-rich conditions~\cite{David_Bi4I4}. In this work, we considered distinct chemical conditions as well as the existence of competing phases, as will be described below, to provide insights on the relation between growth conditions and resulting electronic properties.

As can be seen in Eq.~\ref{Eq:defect_form_energy}, the formation energy is a function of the atomic chemical potentials $\mu_i$, representing the relative abundance of Bi and I species in the growth conditions. We evaluated the chemical potentials $\mu_\text{Bi}$ and $\mu_\text{I}$ in distinct limits to consider the different growth conditions. In a Bi-rich condition, $\mu^{rich}_\text{Bi}$ is calculated from the bulk bismuth, which exhibits $P2_1/m$ symmetry. Therefore, from the thermodynamic equilibrium condition, $\mu_\text{I}^{poor} = (\mu_{\text{Bi}_{4}\text{I}_{4}} - 4\mu^{rich}_\text{Bi})/4$, where $\mu_{Bi_{4}I_{4}}$ is the calculated total energy per formula unit. On the other hand, in I-rich conditions, $\mu^{rich}_\text{I}$ is obtained from the bulk iodine, which exhibits $Cmce$ symmetry. In this condition $\mu_\text{Bi}^\text{poor} = (\mu_{\text{Bi}_{4}\text{I}_{4}} - 4\mu^\text{rich}_\text{I})/4$.

\begin{figure}[tp]    
    \includegraphics[scale=0.101]{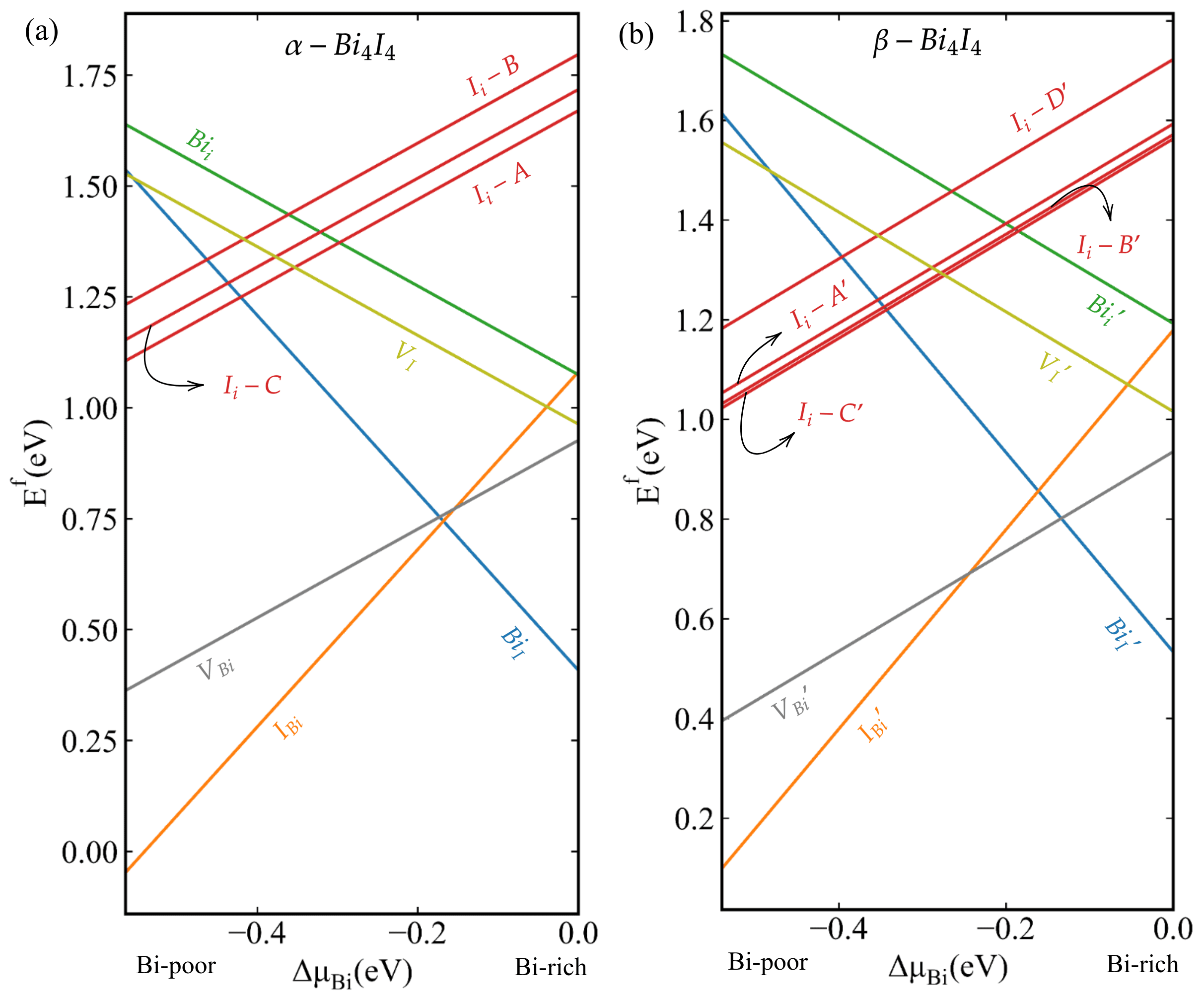}
    \caption{Formation energies of neutral native defects, as a function of chemical potential $\Delta\mu_{Bi}$ for the (a) $\alpha$ and (b) $\beta$ phases. Here, the Kröger-Vink notation is adopted.}
    \label{fig:alphabeta_neutral}
\end{figure}

\begin{figure*}[!ht]    
    \includegraphics[width=1\linewidth]{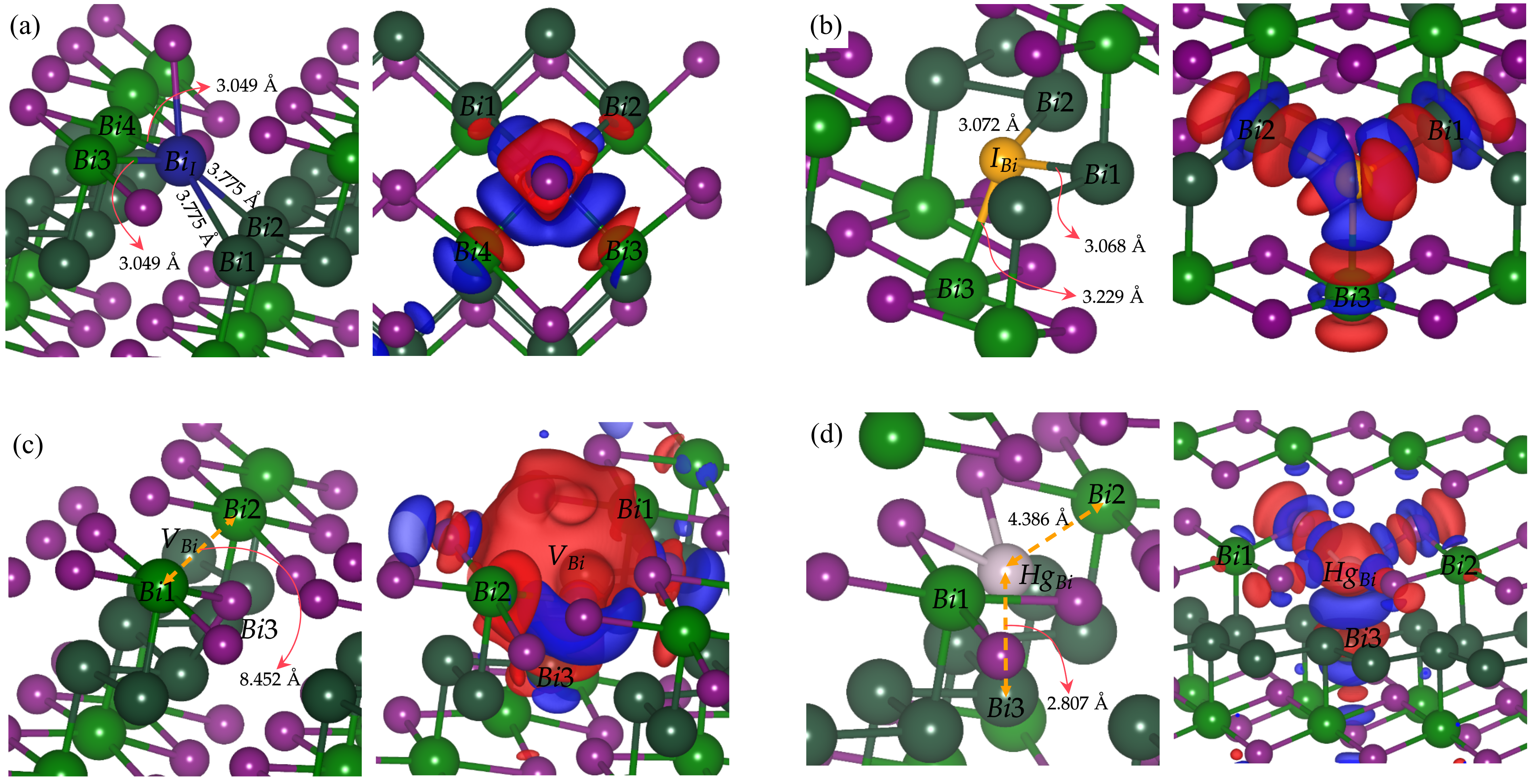}
    \caption{Relaxed structures and charge density difference isosurfaces of the $\alpha$ phase in the presence of native defects and impurities. In the left panels of (a), (b), (c), and (d) we show the obtained equilibrium geometries of $\alpha$-Bi$_4$I$_4$ in the presence of Bi$_\text{I}$, I$_\text{Bi}$, V$_\text{Bi}$, and Hg$_\text{Bi}$, respectively. Light green and dark green represent the outer and inner bismuth atoms, respectively. Iodine atoms are represented in purple. The corresponding charge density differences are shown in the right panels of (a), (b), (c), and (d). The isosurfaces correspond to 0.001 $e$/bohr$^{3}$ charge densities. The red (blue)
    region denotes electron depletion (accumulation).}
    \label{fig:structure_defalpha}
\end{figure*}

\begin{figure*}[!ht]    
    \includegraphics[width=1\linewidth]{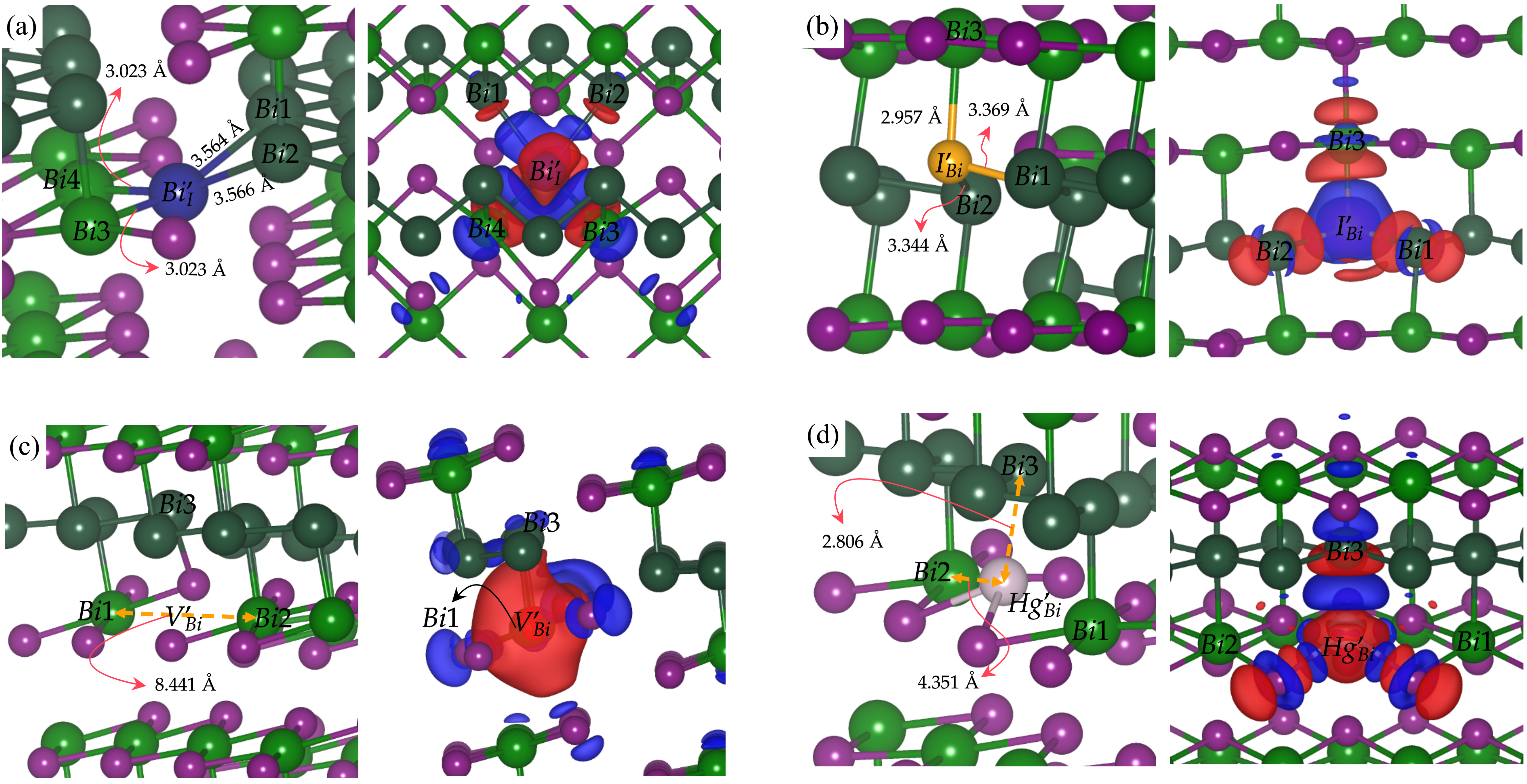}
    \caption{Relaxed structures and charge density difference isosurfaces of the $\beta$ phase in the presence of native defects and impurities. In the left panels of (a), (b), (c), and (d) we show the obtained equilibrium geometries of $\beta$-Bi$_4$I$_4$ in the presence of Bi'$_\text{I}$, I'$_\text{Bi}$, V'$_\text{Bi}$, and Hg'$_\text{Bi}$, respectively. Light green and dark green represent the outer and inner bismuth atoms, respectively. Iodine atoms are represented in purple. The corresponding charge density differences are shown in the right panels of (a), (b), (c), and (d). The isosurfaces correspond to 0.001 $e$/bohr$^{3}$ charge densities. The red (blue)
    region denotes electron depletion (accumulation).}
    \label{fig:structure_defbeta}
\end{figure*}

In table~\ref{tab:formation_energiesNeutral}, we summarize the obtained formation energies for the native defects in both the $\alpha$ and $\beta$ phases for Bi-rich (I-poor) and Bi-poor (I-rich) conditions. Values with an star were reported in our previous work~\cite{David_Bi4I4}. The variation of the formation energies as a function of $\Delta \mu_\text{Bi} =\mu_\text{Bi}^\text{poor}-\mu_\text{Bi}^\text{rich}$ is shown in Fig.~\ref{fig:alphabeta_neutral}.
As expected, we obtain similar formation energies of defects in the $\alpha$ and $\beta$ phases, indicating that the concentration of neutral point defects is roughly independent of the Bi$_4$I$_4$ crystal structure. We also observe that bismuth and iodine interstitials have higher formation energies than the other native defects and are thus expected to appear in small concentrations. More importantly, our findings suggest that the bismuth antisites have low formation energies under Bi-rich conditions, followed by the bismuth and iodine vacancies. The relaxed structures of bismuth antisites in $\alpha$ (Bi$_\text{I}$) and $\beta$ (Bi'$_\text{I}$) phases are shown in Figs.~\ref{fig:structure_defalpha}(a) and~\ref{fig:structure_defbeta}(a), respectively. The surrounding atoms are denoted by Bi$1$, Bi$2$, Bi$3$, and Bi$4$ in our relaxed structures. We find Bi$_\text{I}$-Bi bond lengths of 3.05 and 3.75 \AA{} in the $\alpha$ phase, and of 3.02 and 3.56 \AA{} in the $\beta$ phase. These findings indicate that Bi-antisites tend to modify the interactions between chains in both phases of Bi$_4$I$_4$. In the pristine systems, the Bi$3$-I bond length is around 3.13 (3.11) \AA{} in $\alpha$($\beta$) phase, while Bi$1$-I is around 3.76 (3.84) \AA{} in the $\alpha$($\beta$) phase.
In the case of bismuth vacancies, we do not observe any strong modification in the structures, as shown in Figs.~\ref{fig:structure_defalpha}(c) and ~\ref{fig:structure_defbeta}(c), with Bi$1$-Bi$2$ bond lengths of around 8.4 \AA{} in both phases. In pristine $\alpha$ and $\beta$ phases the corresponding distance is around 8.4 \AA{}.

For comparison, in the case of iodine interstitials in the $\alpha$ phase, the formation energy difference is $\Delta E^{f} = 1.26$ eV concerning the bismuth antisites. Regarding the vacancies, we find that iodine vacancies have higher formation energies than bismuth vacancies. Therefore, our findings suggest that the neutral charge Bi-antisites (Bi$_\text{I}$) and vacancies (V$_\text{Bi}$) are expected to be the dominant defects in Bi$_4$I$_4$ under Bi-rich conditions. We mention that recent scanning tunneling microscopy measurements observed the presence of Bi-vacancies in monolayers of Bi$_4$I$_4$~\cite{Mu2023}.

\begin{table}[!htb]
	\centering
	\caption{Formation energies (eV), E$^{f}$, of neutral native defects in $\alpha$ and $\beta$-Bi$_4$I$_4$ for Bi-rich (I-poor) and Bi-poor (I-rich) conditions. Values with a star were reported previously in Ref. \cite{David_Bi4I4}.} 
	\label{tab:formation_energiesNeutral}
   \begin{ruledtabular}
	\begin{tabular}{cccc}
		\textbf{System} & \textbf{Defect} & \textbf{E$^f$ (eV)} Bi-rich & \textbf{E$^f$ (eV)} Bi-poor  \\ \hline 
		  \multirow{9}{*}{$\alpha$-Bi$_4$I$_4$} & Bi$_\text{I}$ & 0.41$^*$  & 1.54 \\
		                                   & I$_\text{Bi}$ & 1.08$^*$  & -0.05 \\
                                          & Bi$_i$ & 1.07$^*$    & 1.64 \\
                                          & I$_i$-A & 1.67$^*$   & 1.11 \\
                                          & I$_i$-B & 1.80$^*$   & 1.23 \\
                                          & I$_i$-C & 1.72$^*$   & 1.15 \\
                                          & V$_\text{Bi}$ & 0.93$^*$   & 0.36 \\
                                          & V$_\text{I}$ & 0.96$^*$    & 1.53 \\ \hline 
         \multirow{9}{*}{$\beta$-Bi$_4$I$_4$} & Bi'$_\text{I}$ & 0.53$^*$  & 1.61 \\ 
		                                   & I$_\text{Bi}$-A' & 1.18$^*$ & 0.10 \\
                                          & I$_\text{Bi}$-B' & 1.72 & 0.64 \\
                                          & Bi'$_i$ & 1.19$^*$   & 1.73 \\
                                          & I$_i$-A' & 1.59$^*$   & 1.05 \\
                                          & I$_i$-B' & 1.57$^*$   & 1.03 \\
                                          & I$_i$-C' & 1.56$^*$   & 1.02 \\
                                          & I$_i$-D' & 1.72$^*$   & 1.18 \\
                                          & V'$_\text{Bi}$ & 0.94$^*$    & 0.39 \\
                                          & V'$_\text{I}$ & 1.02$^*$  & 1.56 \\ 
 \end{tabular}
\end{ruledtabular}
\end{table}
%%%%-------------------------------------------------------------------

On the other hand, in a bismuth-poor (I-rich) environment we observe a distinct feature. In this limit the iodine antisites (I$_\text{Bi}$ and I$_\text{Bi}-A'$ in Table.~\ref{tab:formation_energiesNeutral}) are preferred due to their low formation energies, as emphasized by $E^{f} = -0.05$ eV ($0.10$ eV) in the $\alpha$ ($\beta$) phase. As a result, our findings suggest that this type of defect should the dominant native defect in Bi$_4$I$_4$ under bismuth-poor conditions, for the chemical potentials considered. The relaxed structures of the $\alpha$ and $\beta$ phases in the presence of this type of point defect, shown in Figs.~\ref{fig:structure_defalpha}(b) and~\ref{fig:structure_defbeta}(b), indicate a weak local distortion of the chains, which have Bi-I equilibrium bond lengths of 3.07 and 3.23 \AA{} in the $\alpha$ phase. These bond lengths are 3.04 \AA{} (Bi$1$-Bi') and 3.07 \AA{} (Bi$3$-Bi') in the pristine phases.
According to our calculations, we find that the formation of bismuth vacancies is more energetically favorable in Bi-poor conditions than in Bi-rich conditions. Surprisingly, however, they exhibit the second lowest formation energy across the entire chemical potential range.
Therefore, without considering competing phases, iodine antisites and bismuth vacancies are expected to be the dominant defects in a bismuth-poor growth condition. 

We also studied the case where the existence of competing phases introduce additional limits to bismuth and iodine chemical potentials. As described in the Supplemental Material, we find that the formation energies within these new limits correspond to essentially a region near the Bi-rich conditions, where the bismuth-antisites and bismuth-vacancies are the dominant defects in both phases of Bi$_4$I$_4$. Therefore, the bismuth antisites and vacancies in neutral charge states are expected to be the dominant defects within this narrow energy window of chemical potentials.

\subsection{Formation energies of Hg impurities in Bi-rich growth conditions}
\label{subsec:form_impuritiesq0}

Motivated by our EDX results on Hg-grown single crystals, we calculate the formation energies of different configurations of Hg impurities in Bi$_4$I$_4$. Using the supercell approximation we have $\alpha$-Bi$_{48}$I$_{48}$Hg$_1$ and  $\beta$-Bi$_{72}$I$_{72}$Hg$_1$ for the interstitial defects, for instance.
The Hg chemical potential was obtained considering the possibility of formation of HgI and HgI$_2$ phases, and according to our calculations give rise to a Hg-rich condition environment. 
Table~\ref{tab:formation_energiesNeutralImpurities} displays the obtained formation energies of Hg impurities in interstitial and substitutional configurations for both the $\alpha$ and $\beta$ phases, where several plausible configurations were considered. In this case, we denote Hg substitutional impurities as Hg$_{Bi}$ (Hg$_{I}$), and the interstitial impurities as Hg$_i$.

\begin{table}[!htb]
	\centering
	\caption{Formation energies (eV), E$^{f}$, of neutral Hg impurities in $\alpha$ and $\beta$-Bi$_4$I$_4$ for Bi-rich (I-poor) conditions.}
	\label{tab:formation_energiesNeutralImpurities}
   \begin{ruledtabular}
	\begin{tabular}{ccc}
		\textbf{System} &  \textbf{Defect} & \textbf{E$^f$ (eV)} Bi-rich  \\ \hline 
		
		  \multirow{4}{*}{$\alpha$-Bi$_4$I$_4$} & Hg$_\text{Bi}$ & 0.47 \\
		                                   & Hg$_\text{I}$ & 0.63  \\
                                          & Hg$_i$-A & 0.98 \\
                                          & Hg$_i$-B &0.84 \\ \hline 

         \multirow{4}{*}{$\beta$-Bi$_4$I$_4$} & Hg'$_\text{Bi}$ &0.47\\
		                                   & Hg'$_\text{I}$ & 0.71 \\
                                          & Hg$_i$-A' & 0.63 \\
                                          & Hg$_i$-B' & 0.76 \\
	\end{tabular}
    \end{ruledtabular}
\end{table}

As can be seen, Hg impurities are energetically favorable to be found at Bi substitutional sites in both $\alpha$ and $\beta$ phases. At the equilibrium geometries of Hg in $\alpha$ phase [Fig.~\ref{fig:structure_defalpha}(d)], we find Hg-Bi$3$ and Hg-Bi$2$ bond lengths of 2.81 and 4.39 \AA{}, respectively. In the $\beta$ phase our calculated relaxed structure [Fig.~\ref{fig:structure_defbeta}(d)] indicates Hg-Bi$3$ and Hg-Bi$2$ bond lengths of 2.81 and 4.35 \AA{}, respectively. The corresponding Bi-Bi distances are around 3.07 and 4.4 \AA{} in the pristine phases. The interstitial configuration Hg$_i$-A(A') formation energies are phase dependent, and are around 0.51 and 0.16 eV higher than the configuration corresponding to Hg at the bismuth sites in the $\alpha$ and $\beta$ phases, respectively. The Hg configuration in iodine sites has a formation energy of 0.16 eV (0.24 eV) higher than the Hg$_{\text{Bi}}$ in the $\alpha$ ($\beta$) phase. Thus, our findings indicate that Hg impurities are mostly likely to be found as substitutional impurities at the Bi sites under the Bi-rich condition, as expected due to the similar ionic radius of Hg and Bi. Regarding the structural properties, Hg antisites do not lead to strong local distortions of the chains in both phases.

\subsection{Electronic structure of Bi$_4$I$_4$ in the presence of native defects and Hg impurities}
\label{subsec:elec_defectsq0}

The different levels of intrinsic $p$ and $n$-type doping in Bi$_4$I$_4$ are ruled by the relative concentration of native defects and impurities in equilibrium conditions, which can also give rise to additional defect/impurity electronic states, change in charge carrier density and mobility, and to the modulation of the band-edges of Bi$_4$I$_4$, and as a result impact the resistivity of Bi$_4$I$_4$. In the work of Chen \textit{et al.}~\cite{PRM_Chen2018}, it was found that most of their $\alpha$-Bi$_4$I$_4$ samples exhibit intrinsic $p$-type character, whereas a small number of samples exhibit dominant $n$-type carriers. According to their quantum transport measurements, in both types of samples, the transport behavior comes from bulk contributions. A small concentration of intrinsic $n$-type carriers was also observed in photoemission experiments~\cite{autes_novel_2016} on $\beta$-Bi$_4$I$_4$. Motivated by those findings and by the variation of the temperature dependence of the electrical resistivity (shown in Fig.~\ref{fig:RhovsT}), we calculate the electronic structure and the local charge density redistribution of both $\alpha$ and $\beta$ phases in the presence of preferred native defects and Hg impurities. We emphasize here that the DFT(PBE) band gap was used in our analysis; however, it is overestimated in comparison with the ones obtained with the modified Becke-Johnson exchange (mBJ) ~\cite{Noguchi2019,Noguchi2021} or Heyd–Scuseria–Ernzerhof (HSE) functionals~\cite{liu2016weak}.

\begin{figure*}[!ht]    
    \includegraphics[scale=0.13]{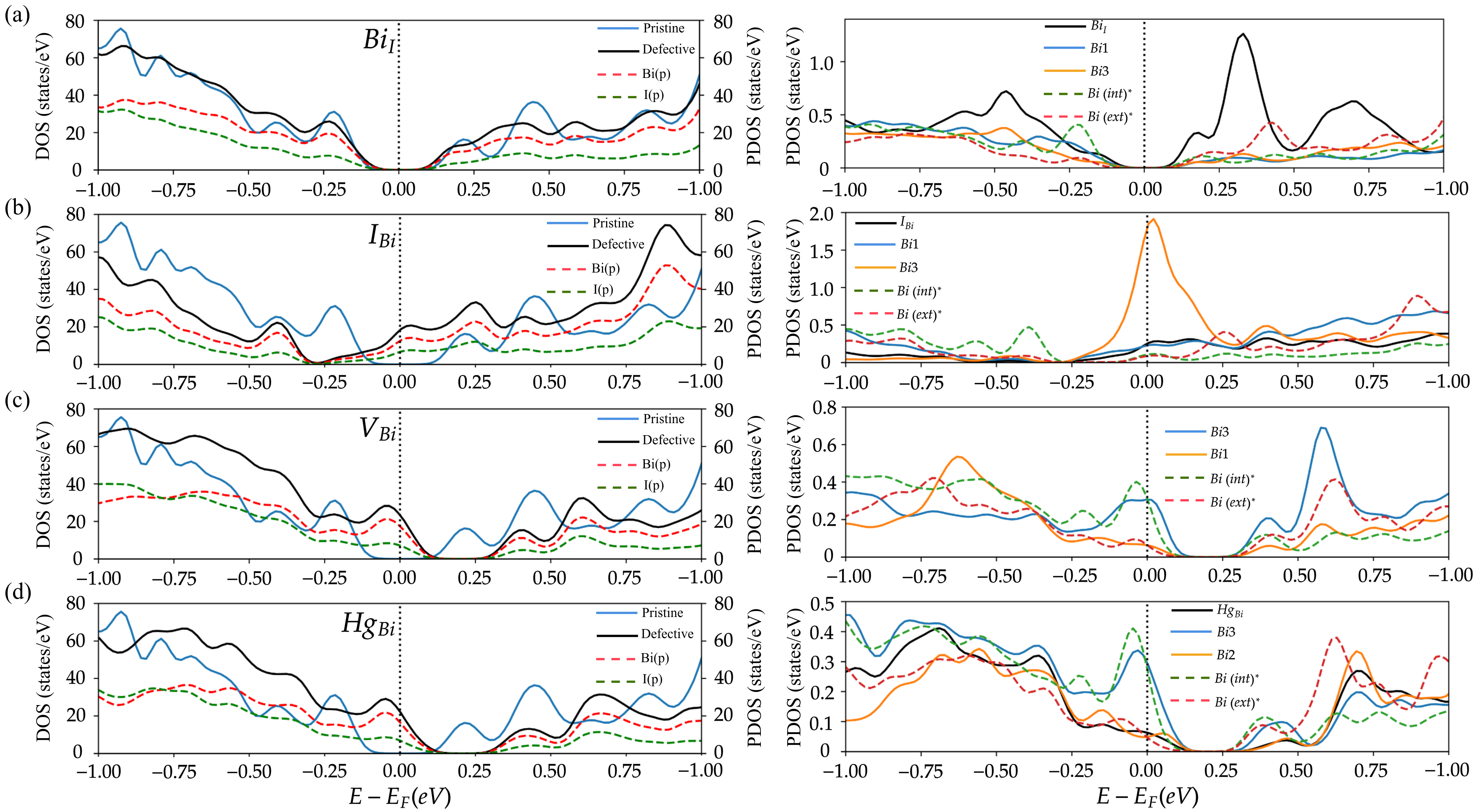}
    \caption{Total and projected density of states of: (a) bismuth antisite, (b) iodine antisite, (c) bismuth vacancy, and (d) Hg-antisite in $\alpha$-Bi$_4$I$_4$. The total density of states of the pristine and defective systems are shown in blue and black (left panels), respectively. The projections on the Bi($p$) and I($p$) orbitals are shown in dashed red and green lines (left panels), respectively. The projected densities of states are obtained for the atoms shown in Fig.~\ref{fig:structure_defalpha} (right panels).}
    \label{fig:dospdos_alfadef}
\end{figure*}

\begin{figure*}[!ht]    
   \includegraphics[scale=0.13]{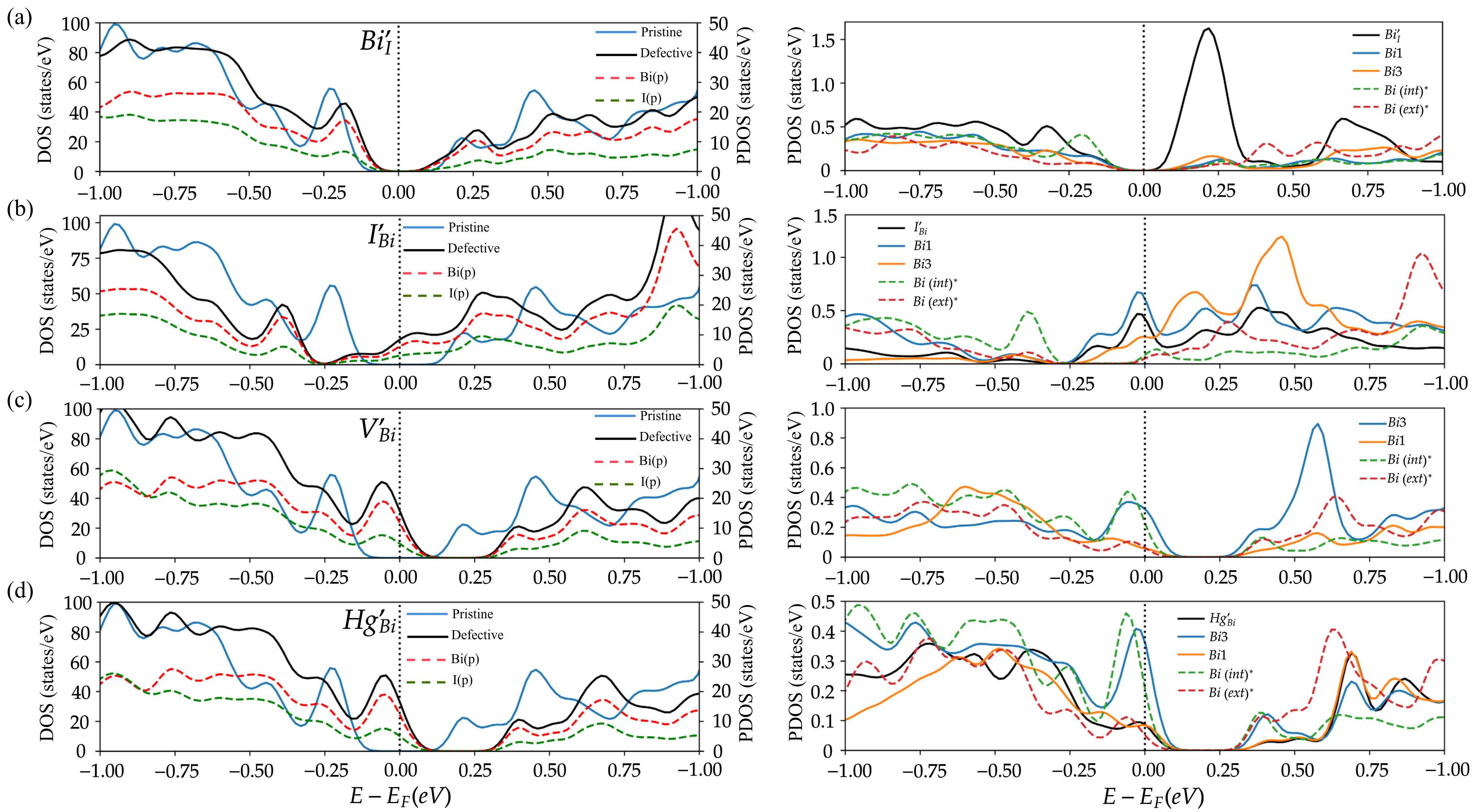}
    \caption{Total and projected density of states of: (a) bismuth antisite, (b) iodine antisite, (c) bismuth vacancy, and (d) Hg-antisite in $\beta$-Bi$_4$I$_4$. The total density of states of the pristine and defective systems are shown in blue and black (left panels), respectively. The projections on the Bi($p$) and I($p$) orbitals are shown in dashed red and green lines (left panels), respectively. The projected densities of states are obtained for the atoms shown in Fig.~\ref{fig:structure_defbeta} (right panels).}
    \label{fig:dospdos_betadef}
\end{figure*}

In Figs.~\ref{fig:dospdos_alfadef} and \ref{fig:dospdos_betadef} we show the obtained density of states (DOS) and projected density of states (PDOS) for $\alpha$ and $\beta$ phases in the presence of (a) bismuth and (b) iodine antisites; (c) bismuth vacancies, and (d) Hg antisites. Here, we also used the same labeling for the surrounding atoms, namely, Bi$1$ and Bi$3$. The less perturbed Bi atoms were denoted as Bi(int) and Bi(ext). For comparison, we show the pristine bulk DOS (blue lines) in these figures. The red and green dashed lines represent an averaged PDOS of Bi-$p$ and I-$p$ states. Our findings indicate that Bi-$p$ states dominate the Fermi energy's electronic states.

More importantly, we find that bismuth antisites do not lead to a strong modification of the electronic density of states around the Fermi energy (black lines), though, in the $\beta$-phase it modulates the band-edges inducing a small reduction of the band gap. They also do not dope Bi$_4$I$_4$, as indicated by the negligible change in the Fermi energy. 
The corresponding charge redistribution observed in the obtained charge density difference indicates that Bi antisites lead to the accumulation of electrons between Bi$_I$ and Bi$1$ and Bi$2$ atoms, as as shown in the right panels of Figs.~\ref{fig:structure_defalpha}(a) and \ref{fig:structure_defbeta}(a), respectively. This charge accumulation is reminiscent of a new interaction between the quasi-1D chains. As shown in the corresponding PDOS, the Bi antisites give rise to resonant electronic states in the valence and conduction band-edges, which may affect the scattering and mobility of charge carriers in Bi$_4$I$_4$. The importance of carriers scattering due to resonant states was emphasized in early studies of Sankey and co-workers~\cite{sankeyAPL} who studied the electronic scattering in semiconductors with zinc-blende structure. We also mention that charge neutral defects, such as the Bi antisites, can induce carriers scattering due to modifications of the local potential due to structural distortions, as early emphasized in Ref.~\onlinecite{gatalskaya1986mechanism}.
One can also note that the Bi$_I$-$p$ electronic states hybridize with the surrounding iodine and bismuth $p$ states, leading to a considerable contribution to of Bi$_I$ to the bottom of the conduction band, as can be seen by induced PDOS peaks in Figs.~\ref{fig:dospdos_alfadef}(a) and ~\ref{fig:dospdos_betadef}(a) (right panels).

Moreover, we observe that the presence of iodine antisites leads to a $n$-type doping of both the $\alpha$ and $\beta$ phases, as can be seen in Figs. ~\ref{fig:dospdos_alfadef}(b) and \ref{fig:dospdos_betadef}(b), respectively. As can be noticed in the obtained DOS, the Fermi level is located above a region of zero DOS (around -0.25 eV) and indicate that iodine antisites act as donors. In comparison with Bi antisites, the iodine antisites leads a considerable perturbation of the electronic structure of Bi$_4$I$_4$, for the considered concentrations associated with our supercell approximation. The largest contribution to the semi-occupied electronic states are due to Bi-$p$ states, from neighbor Bi atoms (Bi$3$ and Bi$1$) and from I antisites, as can be seem from the calculated projected density of states. The calculated charge density differences show an accumulation of electrons around the iodine antisites, due to its largest electronegativity and a reduction of electrons between I antisites and neighbor Bi atoms. On the other hand, a $p$-type doping is observed for bismuth vacancies, as can be seen in Figs.~\ref{fig:dospdos_alfadef}(c) and ~\ref{fig:dospdos_betadef}(c). In these cases, the semi-occupied electronic states are mainly due to Bi$3$-$p$ and Bi(int)-$p$ states near the vacancies. The calculated charge density difference indicate an accumulation of electrons at sites around the vacancies in both phases (as shown by the blue isosurface). Iodine antisites and vacancies introduce semi-occupied states and are expected to act like new scattering centers in Bi$_4$I$_4$, also affecting the resistivity during the resistivity measurements.

Finally, we show in Figs.~\ref{fig:dospdos_alfadef}(d) and ~\ref{fig:dospdos_betadef}(d) the obtained density of states for Bi$_4$I$_4$ in the presence of Hg antisites. In this case we find  a $p$-doping of Bi$_4$I$_4$, as indicated by the downshift of the Fermi level. The semi-occupied states come mostly from Bi$3$ and Bi(int)-$p$ states. These defects induce a depletion of electrons at Hg sites and an accumulation of electrons between Hg and neighbor Bi atoms.
Thus, Hg antisites lead to the $p$-type doping of both phases of Bi$_4$I$_4$ and to semi-occupied states, which are expected to affect the transport properties.

We summarize the effects of the native defects and Hg antisites in the electronic structure of Bi$_4$I$_4$ in Table~\ref{tab:DopyingType}. Overall, our findings indicate that different sorts of defects can give rise to the appearance of different types of doping, semi-occupied electronic states, resonant-states near the valence and conduction band-edges, and a small reduction of the band-gap of Bi$_4$I$_4$. In particular, we find that iodine antisites, bismuth vacancies, and Hg impurities can induce the modification of carrier concentration due to charge doping. We expected that these effects can partially explain the electrical resistivity behavior observed for distinct samples of Bi$_4$I$_4$. We mention that in a recent theoretical study~\cite{CarvaPRB}, the authors found that Mn impurities have important effects on the resistivity of Bi$_2$Te$_3$, mostly associated with the appearance of impurity states near the Fermi energy, their relative energy position, and the modification of charge carrier concentration. As shown by our calculations these features can be induced by bismuth and iodine antisites, as well as by bismuth vacancies.

We emphasize that these findings are in agreement with the tendency displayed by our resistivity measurements, where the increase in annealing temperature leads to the increase of defect/impurity scattering and to the modification of carrier concentration, which are expected to be the dominant mechanism at low-temperatures.
We stress that the quantitative calculations of the electronic scattering due to defects and impurities and its connection with Bi$_4$I$_4$ resistivity is beyond the scope of this work. From the experimental side it also requires careful future investigation relating controlled growth conditions with the resulting charge carriers type and crystal quality. The latter investigations are challenging due to the size of the needle-shape Bi$_4$I$_4$ crystals, making thin films an interesting platform to be explored for fine-tuning the electronic properties of this compound.

\begin{table}[!htb]
	\centering
	\caption{Resulting doping type and modulation of band-gap for the most stable point defects on both Bi$_4$I$_4$ phases.} 
    \label{tab:DopyingType}
   \begin{ruledtabular}
	\begin{tabular}{cccc}
		\textbf{System} &  \textbf{Defect} & Carrier doping & Band gap    \\ \hline 
		
		  \multirow{6}{*}{$\alpha$-Bi$_4$I$_4$} & Bi$_\text{I}$ & None & Reduced \\
		                                   & I$_\text{Bi}$ &  n-type & Closed  \\
                                          & V$_\text{Bi}$ & p-type & Closed \\
                                          & Hg$_\text{Bi}$ & p-type & Closed\\
                                          \hline 

         \multirow{6}{*}{$\beta$-Bi$_4$I$_4$} & Bi'$_\text{I}$ & None & Reduced \\
		                                   & I'$_\text{Bi}$ & n-type & Closed \\
                                          & V'$_\text{Bi}$ & p-type & Closed \\
                                          & Hg'$_\text{Bi}$& p-type & Closed \\
    \end{tabular}
    \end{ruledtabular}
\end{table}

\section{Conclusions}
\label{sec:conc}

We have studied the energetics and effects of native defects and Hg impurities on the electronic properties of Bi$_4$I$_4$. We observe that the change of the electrical resistivity of our Bi$_4$I$_4$ samples, corresponding to the $\alpha$ phase, can be ascribed by the formation of defects and impurities. By means of DFT calculations we addressed the formation energies and the role of native point defects and Hg impurities on the electronic properties of both the $\alpha$ and $\beta$ phases. Regarding the native defects, our findings suggest that bismuth antisites, iodine antisites and bismuth vacancies are the energetically favorable native defects in Bi$_4$I$_4$. The calculated electronic structure indicate that bismuth antisites do not dope the system and leads to a tiny modulation of Bi$_4$I$_4$ band-gap and resonant-states in the band-edges. Iodine antisites induce $n$-type doping while and bismuth vacancies to $p$-type doping, with the presence of semi-occupied electronic states. Our calculations also suggest that a $p$-type doping is induced by Hg antisites, with semi-occupied states as well. We expect that these defects and impurities will modify the mobility, carrier density, scattering, and the electrical resistivity of Bi$_4$I$_4$, partially explaining the distinct resistivity profiles of Bi$_4$I$_4$ samples.

\begin{acknowledgments}
W.H.B. and G.H.C. acknowledge FAPEMIG, CNPq (in particular Grant 402919/2021-1), and the National Laboratory for Scientific Computing (LNCC/MCTI, Brazil) for providing HPC resources of the SDumont supercomputer, which have contributed to the research results, URL: http://sdumont.lncc.br. M.C.O.A. acknowledges FAPEMIG, CNPq and CAPES. V.M. and C.D.H. acknowledge FAPESP grants 2018/19420-3, 22/00992-2, 23/11836-4. J. L. J. acknowledges FAPESP JP grant 2018/08845-3 and CNPq Grant 
No. 31005/2021-6. The work at the University of Warwick was supported by EPSRC, UK, through Grant EP/T005963/1.
\end{acknowledgments}

\bibliography{main}

\end{document}